# Wafer-Scale Epitaxy of Flexible Nitride Films with Superior Plasmonic and Superconducting Performance

Ruyi Zhang,[†,‡] Xinyan Li,[§] Fanqi Meng,[§] Jiachang Bi,[†,‡] Shunda Zhang,[†,‡] Shaoqin Peng,[†,‡] Jie Sun,[†] Xinming Wang,[†] Liang Wu,[∥] Junxi Duan,[⊥] Hongtao Cao,[†] Qinghua Zhang,[*,§] Lin Gu,[§] Liang-Feng Huang,[*,†] and Yanwei Cao[*,†,‡]

[†] Ningbo Institute of Materials Technology and Engineering, Chinese Academy of Sciences, Ningbo 315201, China
[‡] Center of Materials Science and Optoelectronics Engineering, University of Chinese Academy of Sciences, Beijing 100049, China
[§] Beijing National Laboratory for Condensed Matter Physics, Institute of Physics, Chinese Academy of Sciences, Beijing 100190, China
[∥] Faculty of Materials Science and Engineering, Kunming University of Science and Technology, Kunming 650093, China
[⊥] School of Physics, Beijing Institute of Technology, Beijing 100081, China

**ABSTRACT:** Transition-metal nitrides (e.g., TiN, ZrN, TaN) are incredible materials with excellent complementary-metal-oxide-semiconductor compatibility and remarkable performance in refractory plasmonics and superconducting quantum electronics. Epitaxial growth of flexible transition-metal nitride films, especially at wafer-scale, is fundamentally important for developing high-performance flexible photonics and superconducting electronics, but the study is rare thus far. This work reports the high-quality epitaxy of 2-inch titanium nitride (TiN) films on flexible fluorophlogopite-mica (F-mica) substrates via reactive magnetron sputtering. Combined measurements of spectroscopic ellipsometer and electrical transport reveal the superior plasmonic and superconducting performance of TiN/F-mica films owing to the high single crystallinity. More interestingly, the superconductivity of these flexible TiN films can be manipulated by the bending states, and enhanced superconducting critical temperature $T_C$ is observed in convex TiN films with in-plane tensile strain. Density-functional-theory calculations uncover that the strain can tune the electron-phonon interaction strength and resultant superconductivity of TiN films. This study provides a promising route towards integrating scalable single-crystalline conductive transition-metal nitride films with flexible electronics for high-performance plasmonics and superconducting electronics.

# INTRODUCTION

The rapid development of flexible functional materials has been extensively promoting the booming of smart, lightweight, and wearable electronics,[1-3] in which the physical properties (e.g., resistivity, magnetism, ferroelectricity/piezoelectricity) can be tuned by strain for tailored functionalities.[4-8] In general, the difficulty of integrating inorganic materials (fragile but multifunctional) with flexible electronics can be overcome by depositing their films on flexible substrates. Ideally, single-crystalline films are more preferred over other less-crystalline films due to superior physical performance (e.g., piezoelectric coefficient, carrier mobility, ferromagnetic resonance width).[9-11] However, the epitaxial growth of films on the flexible substrates is still very challenging for many functional materials, especially at wafer-scale, a prerequisite for massive device fabrications. So far, only small groups of flexible single-crystalline films, including two-dimensional materials (e.g., graphene, $MoS_2$, h-BN),[12-15] a few transition-metal oxide (TMO) films,[16-21] etc., have been reported.

Recent years have also witnessed the rise of flexible conductive materials with high plasmonic and superconducting performance, which can be exploited for flexible photonics and superconducting electronics.[22-27] Among these materials, transition-metal nitrides (TMNs, e.g., TiN, ZrN, TaN) have recently attracted extensive research attention owing to their excellent complementary-metal-oxide-semiconductor (CMOS) compatibility and remarkable performances in both refractory plasmonics and superconducting electronics.[28-30] As an archetype among these TMNs, TiN plays a continuously rising role in refractory plasmonics and superconducting electronics, which is beyond its traditional roles of decorations, hard coatings, and CMOS migration barriers. On one hand, due to extraordinary plasmonic performance and endurance in extreme environments (e.g., high temperature and intensive illumination), refractory TiN films are very promising in many novel photonic applications, such as sensors, energy harvesting, heat-assisted magnetic recording, and to name a few.[31-34] On the other hand, superconducting TiN (critical temperature $T_C \sim 5.6$ K in bulk) possesses one of the highest kinetic inductance,[35] which makes it a very attractive material in many superconducting devices, including single-photon detector, superconducting phase shifter, quantum resonators, and qubits.[36-37] Wafer-scale epitaxial TMN films (e.g., TiN films) have been extensively grown on rigid wafers, such as Si and $Al_2O_3$.[38-39] However, the epitaxy of flexible TMN films is very rare, especially at wafer-scale. Moreover, the study of the bending effect on the physical properties (e.g., superconductivity) of conductive TMN films is fairly lacking, which is fundamentally important for both physics and applications in flexible electronics.

In this work, the wafer-scale epitaxy of high-quality flexible TiN films on 2-inch fluorophlogopite-mica (F-mica) wafers is developed by a scalable technique of reactive magnetron sputtering. Owing to the high single crystallinity, these flexible TiN films demonstrate superior plasmonic and superconducting performance, which can compete with those deposited on rigid substrates but far exceed those grown on flexible organic substrates. More interestingly, the flexible

TiN films show strain-tunable superconductivity upon various bending, which results from strain-dependent electron-phonon interaction strength in TiN films through the density-functional-theory (DFT) calculations. This study provides a very promising route towards scalable flexible conductive epitaxial nitride films for high-performance flexible refractory plasmonic and superconducting electronics.

**RESULTS AND DISCUSSION**

The wafer-scale fabrication of 120 nm-thick flexible TiN films is schematically illustrated in Figure 1a. As seen, epitaxial TiN films were directly grown on 2-inch F-mica wafers via a homemade reactive magnetron sputtering using a 2-inch pure titanium (99.999%) target and pure nitrogen (99.999%) reactive gas. These TiN films become flexible after mechanical exfoliation of F-mica substrates. F-mica [$KMg_3Al(Si_3O_{10})F_2$] is a synthetic two-dimensional van der Waals layered material with lattice parameters $a = 5.308$ Å, $b = 9.183$ Å, $c = 10.139$ Å, $\alpha = 90°, \beta = 100.07°, \gamma = 90°$ (Figure 1b).[40] F-mica was carefully selected for the following reasons. Firstly, F-mica is chemically inert, thermally stable up to ~1100 °C, and atomically flat on the surface.[18, 41] These properties are highly favorable for refractory plasmonic applications and patterning devices on F-mica. Secondly, F-mica is highly insulating with low dielectric loss in the high-frequency range, which is essential for superconducting quantum resonators and qubits with long coherence time.[42] Lastly, F-mica is mechanically robust, which can be exfoliated to thicknesses of micrometers and repeatedly bent.[17-21, 43] Generally, the muscovite mica can be stable below the growth temperature of 700 °C.[44] The thermally more stable F-mica were used to synthesize highly-crystalline topological insulator,[45] metal,[46] and functional oxide[11, 18, 47] films via different deposition techniques (Table S1 in the Supporting Information). For example, single-crystalline $Bi_2Se_3$ films have been successfully grown on F-mica substrates using a vapor-phase deposition.[45] Atomically flat Pt/F-mica epitaxial films with high crystallinity and thermal stability could be deposited by sputtering.[46] The high-quality functional oxide films (such as $CuFe_2O_4$,[11] $Pr_{0.5}Ca_{0.5}MnO_3$,[18] and Mn-doped $Na_{0.5}Bi_{0.5}TiO_3$-$BaTiO_3$-$BiFeO_3$[47]) on F-mica substrates were grown by pulsed laser deposition[11, 18] and chemical solution deposition.[47] Despite the films mentioned above, very few reported the epitaxy of rock-salt TMN films on F-mica. Figure 1b illustrates the epitaxial relationship between the TiN films and F-mica substrates. Similar to the cubic TMO counterparts,[19-20] TiN/F-mica films can follow the epitaxial relationship expressed as, TiN[11-2] ∥F-mica[100] and TiN[1-10] ∥F-mica[010] for the in-plane directions and TiN(111) ∥F-mica(001) for the out-of-plane direction.

Figure 1c-f show the photo image and crystal structures of 2-inch TiN/F-mica films. As seen in Figure 1c, the 2-inch TiN/F-mica film with a thickness of 120 nm is highly homogeneous with bright golden luster, same as those grown on rigid substrates.[38, 48-49] The wide-range $2\theta$-$\omega$ scan in Figure 1d reveals a preferred (111) orientation for the TiN film on (001)-oriented F-mica without detectable secondary phases. The rocking curve measured around TiN(111) diffraction in Figure 1e

shows a full width at half maximum (FWHM) of 0.32°, indicating a high crystallinity. The surface morphology was scanned by atomic force microscopy (AFM) (Figure S1a in the Supporting Information), showing a smooth surface with root-mean-square roughness (RMS) of 1.7 nm. To further understand the crystal symmetry of TiN lattice, $\phi$ scans were collected around TiN (002) and F-mica (202) diffractions. As seen in Figure S1b, it is 3-fold symmetry for F-mica and 6-fold symmetry for TiN films. Among the six diffraction peaks in TiN $\phi$ scan, three peaks share the same $\phi$ angles as those in F-mica, whereas the other three peaks show azimuthal 60° rotations with respect to F-mica. It is noted that an analogous 6-fold symmetry has been observed in the $\phi$ scan of TiN film grown on $Al_2O_3$ (0001) substrate (Figure S2 in the Supporting Information) using the same growth parameters and $\phi$ scans of TiN/$Al_2O_3$ (0001) films in previous reports.[50] The 6-fold symmetry rather than the 3-fold symmetry shown in TiN $\phi$ scans on these substrates suggests that these TiN films possess grains with a twin structure,[49] which had also been seen in cubic oxide films grown on mica in previous studies.[19-20] Then, as expected from Figure 1b, the epitaxial relationship between TiN film and F-mica (despite the twin structure) can be determined. The reciprocal space mappings (RSMs) around the F-mica (-207) and TiN(113) diffractions in Figure 1f further confirm the epitaxial relationship. The (111) and (11-2) layer spacings extracted from RSMs are 2.448 Å and 1.731 Å, respectively. According to these layer spacings, the TiN/F-mica film is cubic with a lattice parameter ~ 4.240 Å, very close to the lattice parameter 4.244 Å of stoichiometric TiN.[51] It is worth noting that the epitaxial method of TiN films on F-mica demonstrated here is applicable to other transition-metal nitride films (e.g., NbN, ZrN, TaN), which hold the same rock-salt structure and close lattice parameters (4.2 ~ 4.4 Å).[52] As seen in Figure S3 in the Supporting Information, comparing to the TiN films on F-mica, the TiN films deposited on both flexible polyimide and copper foil (low crystallinity or thermal stability) show poor crystalline quality, indicating that thermally stable single-crystalline F-mica substrate plays an important role in synthesizing high-quality flexible TiN films. The X-ray photoemission spectroscopy (XPS) survey scan in Figure S4 in the Supporting Information shows that all TiN/F-mica films are highly stoichiometric, which is crucial for plasmonic and superconducting properties.

To further investigate the crystal structures of TiN/F-mica epitaxial films at the atomic scale, the transmission electron microscope (TEM) characterization was carried out (Figure 2). As demonstrated in Figure 2a-c, clear single-crystalline patterns can be seen in the selected area electron diffraction (SAED) images recorded at the regions of the TiN film (viewed along TiN[1-10] zone axis), the F-mica substrate (viewed along F-mica[010] zone axis), and the interface between the film and substrate. These SAED patterns suggest TiN/F-mica films follow the epitaxial relationship expressed as, TiN[11-2] ∥F-mica[100] and TiN[1-10] ∥F-mica[010] (in-plane) as well as TiN(111) ∥F-mica(001) (out-of-plane), which agree well with the HRXRD results. The annular bright-field (ABF) image in Figure 2d shows that the interface is atomically sharp and TiN film is of high-crystalline qualities. The atomic arrangement of titanium in TiN can be seen in the high-angle annular dark-field (HAADF) image (Figure 2e). According to the HAADF image, layer spacings of ~ 2.45 Å and ~ 1.73 Å can be obtained for the (111) and (11-2) planes, respectively, in good agreement with the values (~ 2.448 Å and 1.731 Å) extracted from the RSM data.

The collective excitations (plasmons) of conduction electrons in TiN films were investigated by spectroscopic ellipsometry (SE). The dielectric functions were derived from SE data fitted with the well-established Drude-Lorentz model (Figure S5 in the Supporting Information).[38, 48, 53] The dielectric functions of the high-quality epitaxial TiN/MgO* (001) films[54] and flexible TiN/PMMA* films[55] were also adapted from the literature as a reference. As seen in Figure 3a,b, the TiN/F-mica shows the most negative real part $\varepsilon'$ of permittivity and relatively low imaginary part of permittivity $\varepsilon''$, indicating a superior plasmonic performance and relatively low dielectric loss. Both the TiN/F-mica and TiN/Al$_2$O$_3$ films possess a screening plasmon energy $\hbar\omega_p$ of 2.61 eV (corresponding to the crossover wavelength $\lambda_p$ of 476 nm), which are very close to the $\hbar\omega_p$ of 2.65 eV in high-quality TiN films.[56] In general, the reflectivity $R$ can be deduced from the formula[38]

$$R = \frac{(n-1)^2 + k^2}{(n+1)^2 + k^2} \quad (1)$$

where the real ($n$) and imaginary ($k$) parts of the complex refractive index can be derived from the relationship $\tilde{n} = n + ik = \sqrt{\varepsilon' + i\varepsilon''}$. As seen in Figure 3c, the reflectivities of TiN/F-mica, TiN/Al$_2$O$_3$, and TiN/MgO* films all reach ~ 90% at the wavelength of ~ 1000 nm, which are much higher than the value (~ 80%) of TiN/PMMA* films. The plasmonic performance of TiN films can be further evaluated by the figure of merit for surface plasmon polariton (FOM$_{SPP}$) defined as $(\varepsilon')^2/\varepsilon''$. As shown in Figure 3d, TiN/F-mica films possess equal FOM$_{SPP}$ values in the near-infrared (NIR) wavelength as that of high-quality TiN/MgO* films in literature.[54] As a comparison, the plasmonic performance of TiN/Al$_2$O$_3$ films is slightly inferior to TiN/F-mica films. Owing to the high crystallinity, the plasmonic performance of TiN/F-mica films far exceeds that of flexible TiN/PMMA* films.[55] Usually, to sustain SPP with meaningful field enhancements, the FOM$_{SPP}$ value should be higher than 1.[38] The TiN/F-mica film shows FOM$_{SPP}$ > 1 above the wavelength of 515 nm, indicating that TiN/F-mica films can be applied in SPP devices operated in the visible-NIR range. It is worth mentioning that our wafer-scale TiN/F-mica films are highly uniform according to the homogeneity test by performing X-ray reflectivity (XRR) and SE measurements across the wafer (Figure S6 in the Supporting Information). Previously, the extraordinary plasmonic performance and thermal stability in extreme environments have been demonstrated in TiN films grown on rigid substrates.[31, 57-58] As the F-mica is thermally stable at 1100 °C, it is reasonable to expect that TiN/F-mica films can also serve well as a unique flexible plasmonic platform operated in extreme environments, which offers an unparalleled advantage over the TiN films deposited on the flexible organic substrates.

Electrical transport measurements can further verify the plasmonic performance in TiN films since the plasmons are highly connected with their electrical properties.[48] More interestingly, superconductivity, another important electronic state in TiN films, can also be revealed from the electrical transport measurements. As seen in Figure 4a, The room temperature resistivity $\rho^{300\ K}$ is 14.9 μΩ·cm and 17.2 μΩ·cm for TiN/F-mica and TiN/Al$_2$O$_3$ films, respectively, both of which are even lower than the $\rho^{300\ K}$ values ~ 20 μΩ·cm for the high-quality TiN films in literature[37] and in great consistency with the superior plasmonic performance observed in Figure 3. At low temperature, the TiN/F-mica films show a sharp superconducting transition with a $T_C$ around 5.3 K, slightly lower than the $T_C$ of 5.5 K in TiN/Al$_2$O$_3$ reference films. Here, the superconducting critical temperature $T_C$ and superconducting critical field $H_C$ are defined as the temperature and field where the resistivity $\rho$ is half of the normal state resistivity. Generally, the structural defects (e.g., grain boundaries, dislocations, point defects) and oxidation tend to suppress the plasmonic performance and superconductivity in TiN films, leading to a relatively high resistivity and low $T_C$ around 4.0 K in reported TiN films.[35-36, 59] The lower resistivity and higher $T_C$ for TiN films here comparing to those in literature indicate the high-quality TiN films in this work. The residual resistance ratio (*RRR*) is defined as $RRR = R^{300\ K}/R^{10\ K} = \rho^{300\ K}/\rho^{10\ K}$, which is ~ 4.3 for the TiN/F-mica film and ~ 7.4 for the TiN/Al$_2$O$_3$ film, respectively. The *RRR* values measured here for the TiN films (~120 nm) on both F-mica and Al$_2$O$_3$ substrates are consistent with the previously reported values (3.2~7.0) for high-quality TiN films (20~200 nm) on Al$_2$O$_3$,[37] further signifying superior crystalline qualities of TiN films in this work. The magnetic field-dependent resistivity at 1.8 K shown in Figure 4b indicates $H_C$ of 0.35 T for TiN/F-mica film and $H_C$ of 1.34 T for the TiN/Al$_2$O$_3$ film. It is noted that both plasmonic and superconducting properties of TiN films strongly depend on the film thickness (Figure S7 in the Supporting Information). Due to prominent quantum size effect and surface oxidation, the carrier density decreases and the superconductivity is suppressed in ultrathin TiN films.[37, 53] Therefore, the metallicity of TiN films is weakened with decreasing the film thickness, leading to a redshift of crossover wavelength and the decreasing of |ε'| at near-infrared wavelength. On the other hand, the superconducting critical temperature $T_C$ also decreases with decreasing TiN film thickness.

To further investigate the superconductivity in bent TiN films, the flexible TiN/F-mica films (Figure 5a) after mechanical exfoliation (the thickness of F-mica was thinned to 15 μm) was attached to Al molds with different radius of curvature. To effectively measure the strain effect on the superconductivity, the current flows in the directions parallel to the bending curvature with the same piece of flexible TiN/F-mica film during electrical transport measurements (Figure 5b and Figure S8 in the Supporting Information). Figure 5c shows the schematics of bent TiN/F-mica films, which experience in-plane tensile strain under convex bending and in-plane compressive strain under concave bending. As shown, the superconductivity of TiN/F-mica films is robust upon different bending statuses. It is noted that no significant wrinkle or crack was observed in TiN films after repeated bending. Convex bending can benefit $T_C$, whereas concave bending reduces $T_C$ in flexible TiN films. To quantitatively analyze the strain effect on $T_C$, the in-plane strain ($\varepsilon_{IP}$) is deduced from the relationship expressed as,[60]

$$\varepsilon_{IP} = \frac{(t_L+t_S)(1+2\eta+\chi\eta^2)}{2R(1+\eta)(1+\chi\eta)} \quad (2)$$

where $\chi = Y_L/Y_S$, $R$ is bending radius, $Y_L$ and $Y_S$ are Young's modulus of the TiN ($Y_L \approx 600$ GPa) and Mica ($Y_S \approx 200$ GPa),[44, 61] $\eta = t_L/t_S$, $t_L$ and $t_S$ are the thickness of the TiN layer and Mica, respectively. As $\eta \ll 1$ ($t_L = 120$ nm, $t_S = 15\ \mu m$), the strain $\varepsilon_{IP}$ can be simply expressed as $\varepsilon_{IP} = \frac{t_S}{2R}$. The stress $\sigma$ in TiN film can also be deduced from strain by using equation $\sigma = Y_L \varepsilon_{IP}$. Therefore, $\varepsilon_{IP} = 0.125\ \%$, $\sigma = 0.75$ GPa for $R = 6$ mm, and $\varepsilon_{IP} = 0.1875\ \%$, $\sigma = 1.125$ GPa for $R = 4$ mm. The monotonic increase of $T_C$ with increasing $\varepsilon_{IP}$ can be seen in Figure 5c. The compressive strain ($\varepsilon_{IP} < 0$) suppresses $T_C$ with minimum $T_C \sim 5.24$ K at $\varepsilon_{IP} = -0.1875\ \%$, whereas the tensile strain ($\varepsilon_{IP} > 0$) enhances $T_C$ with maximum $T_C \sim 5.33$ K at $\varepsilon_{IP} = 0.1875\ \%$.

To gain more insight into the strain-tunable superconductivity in flexible TiN/F-mica films, we carried out DFT calculations (Figure 6a and Figure S9 in the Supporting Information). The calculated phonon spectra and Eliashberg spectral function ($\alpha^2F$) are shown in Figure 6b,c. From the phononic band dispersions and DOSs (Figure 6b), it can be seen that the lower acoustic modes are mainly contributed by the heavier Ti atom, whereas the three higher optical branches are attributed to the lighter N atom. From the Eliashberg spectral function ($\alpha^2F$) (Figure 6c), it can be seen that both the acoustic modes around the Brillouin boundary (180~340 cm$^{-1}$) and optical modes (540~630 cm$^{-1}$) significantly participate in the electron-phonon interaction process. The frequency-resolved electron-phonon coupling constant $\lambda(\omega)$ can be calculated from the integration of $\alpha^2F(\omega)$ spectra $\lambda(\omega) = 2\int_0^\omega dx \cdot \frac{\alpha^2F(x)}{x}$.[62-64] From the variation of $\lambda(\omega)$ in Figure 6c, it can be estimated that the acoustic modes (or Ti atom) and optical modes (or N atom) contribute to the electron-coupling constant by 73% and 27%, respectively. The calculated $\alpha^2F(\omega)$ and $\lambda(\omega)$ were then used to derive $T_C$ using the parametrized semi-empirical McMillan formula[62-63, 65]

$$T_c = \frac{\omega_{\log}}{1.2} \cdot \exp\left[\frac{-1.04(1+\lambda)}{\lambda(1-0.62\mu^*)-\mu^*}\right] \quad (3)$$

where $\omega_{\log}$ (in the unit of K) is the logarithmic average of the phonon frequency; $\lambda$ is the dimensionless electron-phonon coupling constant ($\lambda$ at the maximum frequency shown in Figure 6c); $\mu^*$ is the dimensionless Coulomb pseudopotential. The parameter $\mu^*$ is set to be 0.159 here to make the theoretical $T_C$ equal to the experimental one at $\varepsilon_{IP}= 0$. As demonstrated in Figure 6d, the experimental and theoretical data agree very well with each other. More details about the DFT calculations are described in Figure S9 and S10 in the Supporting Information. To reveal the underlying mechanism for the strain effect on $T_C$, the parameters ($\omega_{\log}$ and $\lambda$) in McMillan formula are further investigated. On one hand, the change of $\omega_{\log}$ indicates the average frequency shift of the phonon modes, which have effective interaction with electrons. On the other hand, the increase (decrease) of $\lambda$ originates from the enhanced (weakened) electron-phonon interaction strength. From Figure 6e, it can be clearly seen that $\omega_{\log}$ has a monotonic decrease with increasing $\varepsilon_{IP}$, which is reverse to the increasing trend in $T_C$. However, $\lambda$ not only has an increasing trend but also exhibits a similar curve shape as that of $T_C$. Therefore, the increase in $T_C$ by a uniaxial strain in the (111) plane is governed by the enhanced electron-phonon interaction strength.

**CONCLUSIONS**

In summary, high-quality flexible conductive epitaxial TiN films with wafer-scale were synthesized by reactive magnetron sputtering. The crystal structures of TiN films were characterized by XRD, AFM, and STEM, whereas their plasmonic and superconducting properties were measured by SE and electrical transport. The flexible TiN films show superior plasmonic and superconducting performance owing to their high single crystallinity. It is also revealed that the superconductivity can be enhanced in bent TiN films, which is closely connected to the bending-dependent electron-phonon interaction from DFT calculations. Our study provides a very promising route towards scalable flexible conductive nitride films for high-performance flexible refractory plasmonic and superconducting electronics.

**EXPERIMENTAL SECTION**

**Film Growth.** The 120 nm-thick TiN films were directly grown on 2-inch F-mica and c-plane $Al_2O_3$ (0001) wafers via a homemade reactive magnetron sputtering using a 2-inch Ti target (99.999%) and pure nitrogen gas (99.999%).[49, 66] The base pressure of the deposition chamber is ~ $3\times 10^{-8}$ Torr. The deposition pressure was maintained at 0.02 Torr. During growth, the substrate temperature was held at 1000 °C, and the RF generator power was set at 100 W during deposition. To guarantee the uniformity of wafer-scale films, the sample holder was rotating at 5 rpm during growth. After deposition, all films were cooled in 0.02 Torr pure $N_2$ atmosphere following a cooling rate of 25 K/min.

**Structural Characterization.** The high-resolution X-ray diffractometer (HRXRD, Bruker D8 Discovery) using monochromatic Cu K$_{\alpha 1}$ radiation with a wavelength of 1.5406 Å was utilized to record the $2\theta$-$\omega$ scans, rocking curves, $\phi$ scans, reciprocal space mappings (RSMs). The surface morphology of TiN films was scanned by atomic force microscopy (AFM, Bruker Dimension ICON SPM). The transmission electron microscope (TEM) samples of TiN/F-mica films were prepared by using focused ion beam (FIB) milling. Cross-sectional lamellas were thinned down to 100 nm-thick at an accelerating voltage of 30 kV with a decreasing current from the maximum 2.5 nA, followed by fine polish at an accelerating voltage of 2 kV with a small current of 40 pA. The atomic structures were characterized using an ARM 200CF (JEOL, Tokyo, Japan) transmission electron microscope operated at 200 kV and equipped with double spherical aberration (Cs) correctors. The annular bright-field (ABF) and high-angle annular dark-field (HAADF) images were acquired at acceptance angles of 11-22 mrad and 68~260 mrad, respectively. The chemical composition of TiN/F-mica films was detected by X-ray photoemission spectroscopy (Kratos AXIS Supra) at normal incidence after the film surface was etched by Ar ions.

**Optical and Electrical Properties Measurements.** The spectroscopic ellipsometry study was performed at a variable-angle spectroscopic ellipsometer (J. A. Woollam M-2000DI) with wavelengths varying from 400 nm to 1700 nm. The dielectric functions of TiN films were derived from spectroscopic ellipsometry data using the Drude-Lorentz model. The temperature ($T$) dependent- and magnetic field ($B$) dependent-resistivity ($\rho$) of TiN films were measured using a Quantum Design physical properties measurement system (PPMS). The $\rho$-$T$ curves from 6 K to 1.8 K at zero magnetic field were scanned at a cooling rate of 0.2 K/min with 0.02 K/step. TiN/F-mica films after mechanical exfoliation were attached to specially designed Al molds with different curvatures to test bending effects on the superconductivity. All the bending tests were performed on the same piece of flexible TiN/F-mica film. In various bending tests, the current directions were always parallel to the bending curvature (along the [11-2] axis of TiN film).

**Density-Functional-Theory Calculations.** The DFT calculations were carried out using Quantum ESPRESSO code package,[67] where the local-density-approximation functional parametrized by Perdew and Zungeris was used to describe the electronic exchange-correlation interaction.[68] The projector augmented wave[69] method was used to treat the interaction between core and valence electrons. To generate the pseudopotentials of Ti and N atoms, the valence configurations of $3s^23p^63d^24s^14p^1$ and $2s^22p^3$ were used, respectively. The occupation of Ti-4p orbital was considered here because of the unavoidable 4s-4p electronic transfer at Ti atoms in solids,[70] which can be seen in the projected electronic density of states (Figure S10a). The energy cutoffs for the plane-wave expansions of the wave function and charge density are 80 and 800 Ry, respectively. The periodic rectangular unit cell of TiN constructed for the calculation can be seen in Figure S9, where the rectangular-cell axes of $a$, $b$, and $c$ are along the [11-2], [1-10], and [111] directions of the rock-salt unit, and the uniaxial in-plane strain is imposed onto the [11-2] axis. The reciprocal $k$ grid used for the calculations of energy, structure, electronic structure, phonon spectra was 10×6×4, and that for the electron-phonon interaction was 20×12×8. The reciprocal $q$ grid used for the lattice-dynamical matrices was 5×3×2. The electronic-state energy levels were smeared using a Gaussian function with a width of 0.02 Ry in the calculations of electronic structure, phonon spectra, and electron-phonon interaction matrices.

## ASSOCIATED CONTENT

### Supporting Information

Supporting Information is available online.

AFM image and $\phi$ scans of TiN/F-mica thin film, photo image and HRXRD characterizations on TiN/Al$_2$O$_3$ (0001) thin film, photo image and $2\theta$-$\omega$ scans for TiN films grown on polyimide and Cu foil, XPS survey scan, fitting details of SE data, homogeneity test through XRR and SE, thickness-dependent plasmonic and electrical transport properties, image of electrical transport measurement, and DFT calculation details.

## AUTHOR INFORMATION

### Corresponding Author

*E-mail: zqh@iphy.ac.cn, huangliangfeng@nimte.ac.cn, ywcao@nimte.ac.cn

### Author Contributions

R. Zhang and Y. Cao conceived the project. R. Zhang, J. Bi, S. Zhang, S. Peng, J. Sun, X. Wang, L. Wu, and J. Duan prepared the samples and performed the HRXRD, SE, and PPMS measurements. X. Li, F. Meng, Q. Zhang, and L. Gu performed the TEM characterizations and analyses. R. Zhang and H. Cao prepared the fittings and analyses of SE data. L. Huang performed the DFT calculations and analyzed the data. All authors discussed the data and contributed to the manuscript.

### Notes

The authors declare no competing financial interest.

## ACKNOWLEDGMENTS

This work was supported by the National Natural Science Foundation of China (Grant Nos. 11874058, 52025025, 52072400, and U2032126), the National Key Basic Research Program of China (Grant No. 2019YFA0308500), the Pioneer Hundred Talents Program of the Chinese Academy of Sciences, the Beijing Natural Science Foundation (Z190010), the Natural Science Foundation of Zhejiang Province, the Beijing National Laboratory for Condensed Matter Physics, and the Ningbo Science and Technology Bureau (Grant No. 2018B10060). This work was partially supported by the Youth Program of National Natural Science Foundation of China (Grant No. 12004399), China Postdoctoral Science Foundation (Grant No. 2018M642500), Postdoctoral Science Foundation of Zhejiang Province (Grant No. zj20180048), and Natural Science Foundation of Ningbo City (Grant No. 202003N4364).

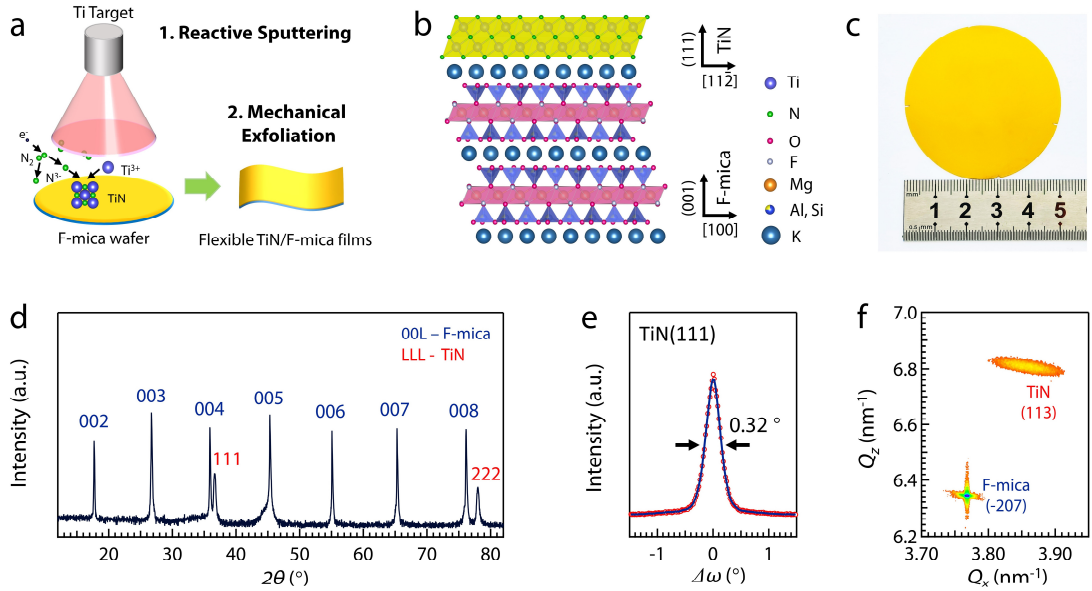

**Figure 1.** Wafer-scale epitaxy, photo image, and HRXRD characterizations of TiN/F-mica films. (a) Schematic of epitaxial TiN films on F-mica wafers using reactive magnetron sputtering. TiN films can be flexible after mechanical exfoliation. (b) The crystal structure of TiN/F-mica films. (c) Photo image of a 2-inch homogenous TiN/F-mica film with a bright golden luster. (d) Wide-range $2\theta$-$\omega$ scan. (e) Rocking curve around TiN (111) diffraction with a FWHM ~ 0.32°. (f) RSM pattern around F-mica (-207) and TiN (113) diffractions.

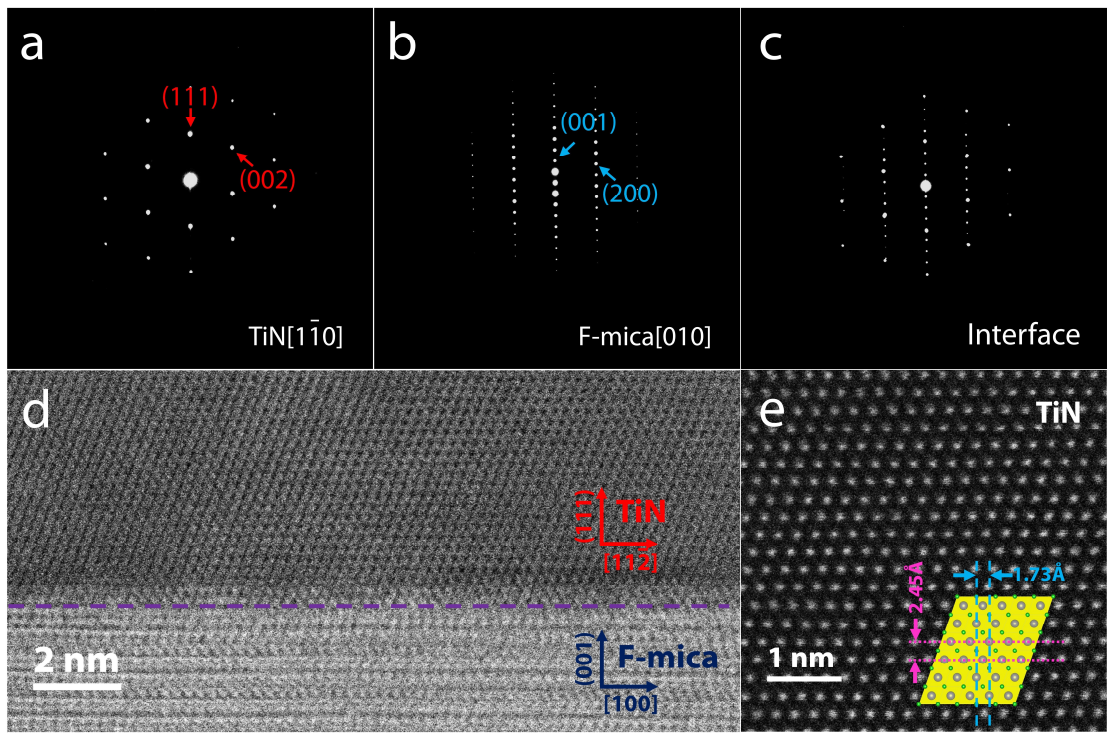

**Figure 2.** TEM characterization of TiN/F-mica films. SAED images were recorded at the regions of (a) the TiN film, (b) the F-mica substrate, and (c) the interface between TiN and F-mica. (d) The ABF image viewed along the TiN [1-10] zone axis. (e) The HAADF image of TiN film.

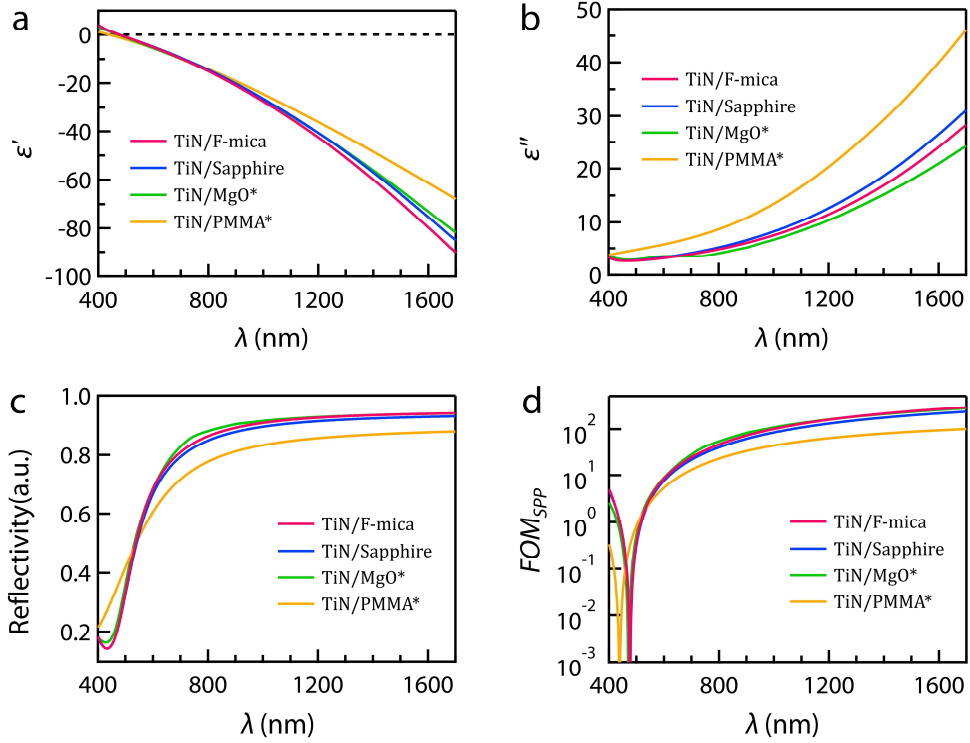

**Figure 3.** Comparisons of plasmonic properties of TiN/F-mica, TiN/Al$_2$O$_3$, TiN/MgO* (001), and TiN/PMMA* films. (a) Real part $\varepsilon'$ and (b) imaginary part $\varepsilon''$ of dielectric functions, (c) extracted reflectivity, and (d) the figure of merit for surface plasmon polariton (FOM$_{SPP}$ = $(\varepsilon')^2/\varepsilon''$) of 120 nm-thick TiN/F-mica films and TiN/Al$_2$O$_3$ reference films in this work, as well as TiN/MgO* (001) and flexible TiN/PMMA* films in literature.[54-55] (Reprinted in part with permission from ref. 54. Copyright 2020 The Optical Society. Reprinted in part with permission from ref. 55. Copyright 2018 American Chemical Society.)

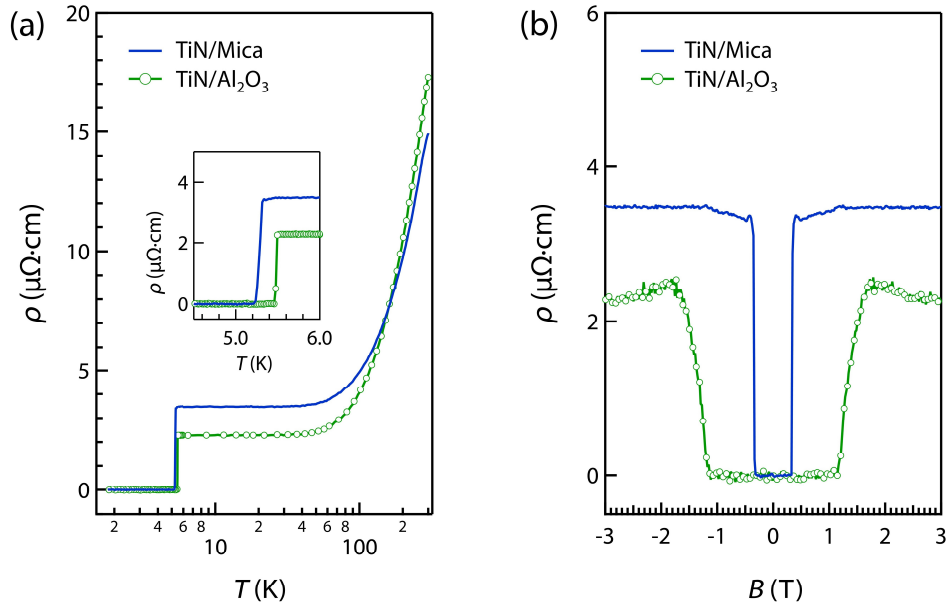

**Figure 4.** Electrical transport properties of TiN/F-mica (at the flat condition) and TiN/Al$_2$O$_3$ films. (a) Temperature ($T$)-dependent resistivity ($\rho$). Inset is the enlarged view near superconducting transition. (b) Magnetic field ($B$)-dependent resistivity ($\rho$) at 1.8 K.

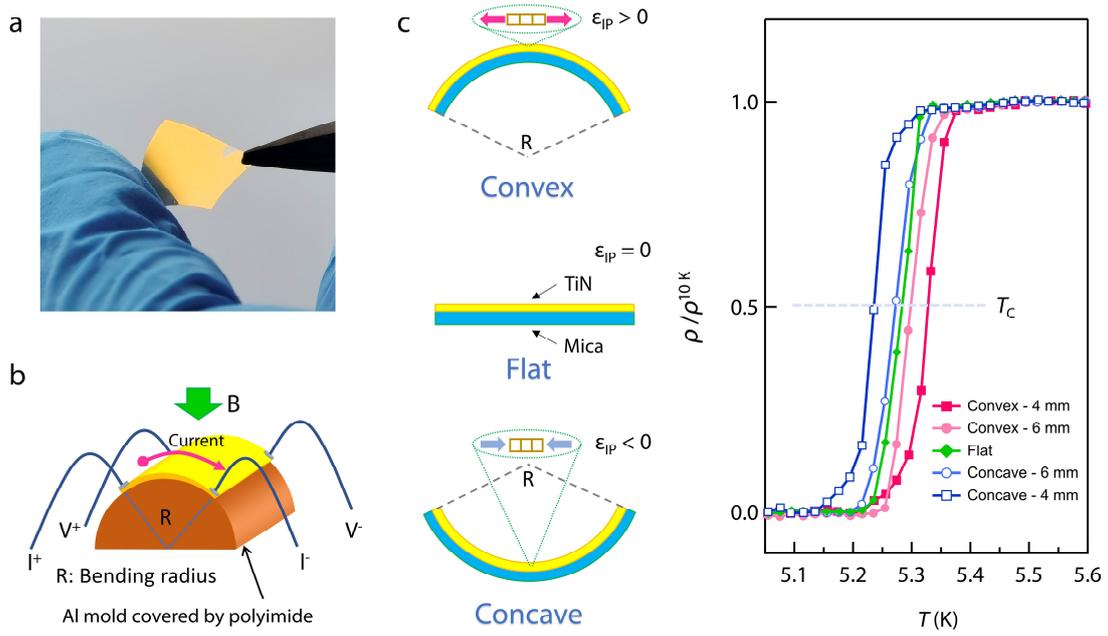

**Figure 5.** Bending curvature-dependent superconductivity of 120 nm-thick TiN/F-mica films. (a) Photo image of a flexible TiN/F-mica film after mechanical exfoliation. (b) Schematic of electrical transport setup. TiN/F-mica films were attached to the Al molds with different curvatures. The current always flows in the direction parallel to the bending curvature. (c) Normalized $\rho$-$T$ curves under different bending statuses.

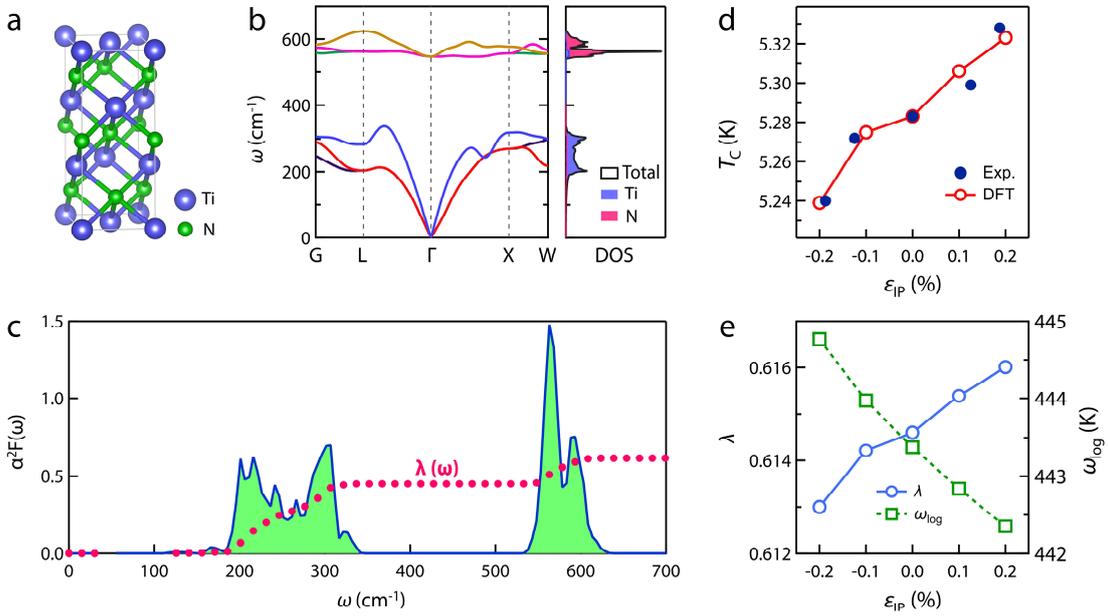

**Figure 6.** Theoretical study of the superconductivity in TiN films. (a) The rectangular unit cell of TiN for DFT calculations. (b) The phonon band dispersions, total phonon DOS, and projected phonon DOS. (c) The Eliashberg spectral function ($\alpha^2F$) and its integration (i.e., electron-phonon coupling constant, $\lambda$). (d) Superconducting critical temperature ($T_C$) versus in-plane strain ($\varepsilon_{IP}$) from experimental data (Exp.) and DFT calculations. (e) In-plane strain-dependent electron-phonon coupling constant ($\lambda$) and logarithmically averaged phonon frequency ($\omega_{log}$).

For Table of Contents Use Only

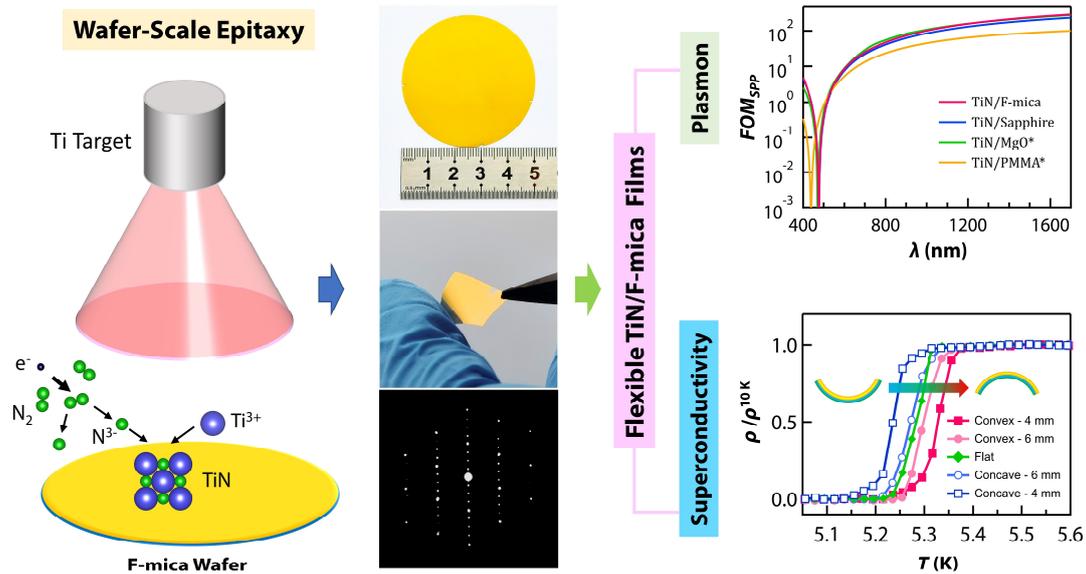